\documentstyle[twoside,fleqn,espcrc2]{article}

\newcommand{\nn}{\nonumber}

\newcommand{\ovl}[1]{\overline{#1}}

\newcommand{\eqn}[1]{(\ref{#1})}

\newcommand{\pslash}{p\kern-1ex /}
\newcommand{\Dslash}{{\cal D}\kern-1.5ex /}

\newcommand{\bpsi}{\overline{\psi}}
\newcommand{\bq}{{\overline{q}}}

\newcommand{\msbar}{{\overline {\rm MS}}}
\newcommand{\vev}[1]{\langle #1 \rangle}

\title{
Perturbative renormalization factors of quark operators for domain-wall QCD
\thanks{Talk presented by Y.~Taniguchi}}

\author{
Sinya Aoki $^{\rm a}$, Taku Izubuchi $^{\rm a}$,
Junichi Noaki
\address{Institute of Physics, University of Tsukuba, Ibaraki 305-8571,
Japan},
Yoshinobu Kuramashi
\address{Department of Physics, Washington University, 
St. Louis, Missouri 63130, USA}
and
Yusuke Taniguchi $^{\rm a}$
}

\begin{document}

\begin{abstract}
We calculate one-loop renormalization factors of several quark
operators including bilinear, three- and four-quark operator
for domain-wall fermion action. Since Green functions 
are constructed for external physical quark fields, our
renormalization method is simple and can be easily applied to
calculation of any quark operators.
Our results show that these renormalized quark
operators preserve several chiral properties of continuum massless
QCD, which can be understood by the property of external quark line
 propagator.
\end{abstract}

\maketitle

% figures here
\makeatletter
\def\setcaption#1{\def\@captype{#1}}
\makeatother

\section{Introduction}

The domain-wall formulation of the massless fermion\cite{Kaplan} was
applied to the lattice QCD (DWQCD) with a simpler form by
Shamir\cite{Shamir93,Shamir95},
anticipating superior features over other quark formulations:
no need of the fine tuning to realize the chiral limit,
no restriction for the number of flavors and no ${\cal O}(a)$ errors.
These features have been proved perturbatively in
\cite{Aoki-Taniguchi,Noaki-Taniguchi}
and results from recent simulations to support existence of the massless
mode in the scaling region\cite{Blum-Soni,Wingate,Blum,Columbia}. 
These advantageous features fascinate us to apply
the domain-wall fermion for 
calculations of weak matrix elements sensitive to
the chiral symmetry such as $B_K$ and other $B$ parameters.
In order to convert several quantities obtained by 
lattice simulations to those defined in some continuum
renormalization scheme(${\it e.g.}, \msbar$),
we must know the renormalization factors\cite{AIKT}.
In this article we summarize our perturbative results of 
renormalization factors for the quark propagator, the bilinear,
three- and four-quark operators consisting of physical quark fields,
together with the perturbative understanding of the chiral properties.

\section{Action and operators}

We adopt the Shamir's action\cite{Shamir93} in this article with
the extra fifth dimensional length set to $N\to\infty$.
In the DWQCD the massless fermion is expressed by the ``physical'' quark
field defined by the boundary fermions 
\begin{eqnarray}
q(n) = P_R \psi(n)_1 + P_L \psi(n)_{N},
\nn \\
\ovl{q}(n) = \bpsi(n)_{N} P_R + \bpsi(n)_1 P_L
\label{eqn:quark}
\end{eqnarray}
with a projection matrix $P_{R/L}=(1\pm\gamma_5)/2$.
We will construct the QCD operators from this quark fields.
The bilinear quark operator is given by
\begin{eqnarray}
{\cal O}_\Gamma = \bq \Gamma q, \quad
\Gamma=1, \gamma_5, \gamma_\mu, \gamma_\mu\gamma_5, \sigma_{\mu\nu}
\end{eqnarray}
and the three-quark operator becomes
\begin{eqnarray}
&&\!\!\!\!\!\!\!\!\!\!\!\!
{\cal O}_{PD} =
 \varepsilon^{abc} 
\left((\bq^C_1)^a \Gamma_X (q_2)^b\right) (\Gamma_Y (q_3)^c),
\nn\\&&\!\!\!\!\!\!\!\!\!\!\!\!
\Gamma_X\otimes\Gamma_Y =
P_{R,L} \otimes P_{R,L},
\end{eqnarray}
where $\bq^C$ is a charge conjugated field and $a,b,c$ are color indices.
The four-quark operator is 
\begin{eqnarray}
{\cal O}_\pm \!\!\!\!\!&=&\!\!\!\!\! \frac{1}{2} \left[
(\bar q_1 \gamma_\mu^L q_2)(\bar q_3 \gamma_\mu^L q_4)
\pm
(\bar q_1 \gamma_\mu^L q_4)(\bar q_3 \gamma_\mu^L q_2) \right],
\nn\\
{\cal O}_1 \!\!\!\!\!&=&\!\!\!\!\!
-C_F (\bar q_1 \gamma_\mu^L q_2)(\bar q_3 \gamma_\mu^R q_4)
\nn\\&&
+ (\bar q_1 T^A \gamma_\mu^L q_2)(\bar q_3 T^A \gamma_\mu^R q_4),
\nn\\
{\cal O}_2 \!\!\!\!\!&=&\!\!\!\!\!
\frac{1}{2N_c} (\bar q_1 \gamma_\mu^L q_2)(\bar q_3 \gamma_\mu^R q_4)
\nn\\&&
+ (\bar q_1 T^A \gamma_\mu^L q_2)(\bar q_3 T^A \gamma_\mu^R q_4),
\nn\\
\gamma_\mu^{R,L} \!\!\!\!\!&=&\!\!\!\!\! \gamma_\mu P_{R,L},
\end{eqnarray}
where $T^A$ is a generator of color $SU(N_c)$ group.

\section{One loop calculation}

We calculate the one loop corrections to the quark propagator and the
quark operators defined in the above.
The point is that our calculation is done in the Green functions
consisting of the ``physical'' quark fields only;
$\vev{q\bq}$, $\vev{{\cal O}_\Gamma q\bq}$,
$\vev{{\cal O}_{PD} q^C_1 \bq_2 \bq_3}$,
and $\vev{{\cal O}_{\pm,1,2} q_1 \bq_2 q_3 \bq_4}$,
where the external quark line plays an important role.
In general the fermion propagator $S_F(p)_{st}$ in the DWQCD has a
complicated form, which connects different five dimensional indices $s,t$.
However the external line propagator, connecting boundary fermion with
some flavor index takes simple form at the physical scale
of the external quark momentum and mass:
\begin{eqnarray}
\vev{q(p)\bpsi(-p)_s} \!\!\!\!\!&=&\!\!\!\!\!
S_q(p) \left[\xi_e L(p)_s-\xi_o R(p)_s \right],
\nn\\
\vev{\psi(p)_s\bq(-p)} \!\!\!\!\!&=&\!\!\!\!\!
\left[R(p)_s\xi_e-L(p)_s\xi_o\right] S_q(p) ,
\label{eqn:barq}
\end{eqnarray}
where $\xi_{e/o}$ are analytic even/odd functions of the momentum and
mass\cite{Noaki-Taniguchi}. The quark propagator in the
continuum,  $S_q$, is given by
\begin{eqnarray}
S_q(p) \!\!\!\!&=&\!\!\!\! \frac{1-w_0^2}{i\pslash + (1-w_0^2) m},
\end{eqnarray}
where $1-w_0^2=M(2-M)$ is a overall factor, which indicates the overlap
between the normalized massless fermion mode and the boundary fermion at
tree level.
The flavor-index dependence only shows up in the factors
\begin{eqnarray}
L(p)_s \!\!\!\!\!&=&\!\!\!\!\!
e^{-\alpha(p)(N-s)}P_R+e^{-\alpha(p)(s-1)}P_L,
\nn\\
R(p)_s \!\!\!\!\!&=&\!\!\!\!\!
e^{-\alpha(p)(s-1)}P_R+e^{-\alpha(p)(N-s)}P_L,
\label{eqn:LRdef}
\end{eqnarray}
where $\alpha(p)$ is an even function of $p$.

With help of \eqn{eqn:barq}, the one loop correction
in the DWQCD can be written in the same form as in the continuum calculation.
For example, the one loop correction to the quark propagator is given by
\begin{eqnarray}
\vev{q(p) \ovl{q}(-p)}_1
= S_q(p) \Sigma_q(p,m) S_q(p),
\label{eqn:one-loop}
\end{eqnarray}
which is same as that in the continuum, except for the overall factor
$1-w_0^2$.
The peculiar feature in the DWQCD is that the Dirac mass $M$ is
renormalized additively and the overall factor $1-w_0^2$ is
shifted by the quantum correction.
Therefore we need renormalization factor $Z_w$ for $w_0$,
which turns out to be
\begin{eqnarray}
&&\!\!\!\!\!\!\!\!\!\!
Z_w = 1+\frac{g^2 C_F}{16\pi^2} z_w .
\end{eqnarray}

By evaluating the quantum corrections $\Sigma_q$ for each Green functions
and summing up all the flavor dependence together with $L_s$ and $R_s$, 
we get renormalization relations for various
quantities.
The bare quark wave function and mass on the lattice are connected 
multiplicatively with the renormalized ones in the $\msbar$ scheme 
at scale $\mu$, by renormalization factors $Z_2$ and $Z_m$,
\begin{eqnarray}
  &&\!\!\!\!\!\!\!\!\!\!
Z_2 (\mu a) = 1+\frac{g^2}{16\pi^2}C_F\left[ 
-\log (\mu a)^2 + z_2 
\right] ,
\label{eqn:wave}
\\&&\!\!\!\!\!\!\!\!\!\!
Z_m (\mu a) = 1+\frac{g^2}{16\pi^2}C_F\left[ 
-3 \log (\mu a)^2 +z_m \right].
\label{eqn:mass}
\end{eqnarray}
The renormalization factor of the bilinear quark operator becomes
\begin{eqnarray}
 &&\!\!\!\!\!\!\!\!\!\!
Z_\Gamma(\mu a) = 1+\frac{g^2 C_F}{16\pi^2}
\left[x_2(\Gamma)\log(\mu a)^2 + z_\Gamma \right] ,
\nn\\&&\!\!\!\!\!\!\!\!\!\!
x_2(\Gamma) = 3(S), 3(P), 0(V), 0(A), -1(T),
\label{eqn:bilinear}
\end{eqnarray}
and the three quark operator is renormalized with
\begin{eqnarray}
&&\!\!\!\!\!\!\!\!\!\!\!
Z_{PD}(\mu a) =
1 + \frac{g^2}{16\pi^2}\Biggl[
\nn\\&&
\left(\frac{3(N_c+1)}{N_c} -\frac{3}{2}C_F\right)\log (\mu a)^2
+ z_{PD} \Biggr].
\end{eqnarray}
The renormalization factors of the four-quark operator is given as
follows for $F=\pm,1,2$.
\begin{eqnarray}
&&\!\!\!\!\!\!\!\!\!\!\!\!
Z_F (\mu a) =
 1 + \frac{g^2}{16\pi^2}\left[
(\delta_F - 2 C_F)\log (\mu a)^2 + z_F \right],
\nn\\
\end{eqnarray}
where $\delta_F$ is a $N_c$ dependent numerical factor\cite{AIKT}.
The finite parts of the renormalization factors
$z_2$, $z_m$, $z_w$, $z_\Gamma$, $z_{PD}$, $z_F$ are given in our
previous paper \cite{AIKT} for various $M$, with and without mean field
improvement.

\section{Chiral properties}

Now we notice the relation
$Z_S=Z_P=Z_m^{-1}$ and $Z_V=Z_A$ \cite{AIKT}, which suggest that the chiral
Ward-Takahashi identity holds exactly.
We can also see that
the three- and four-quark operators can be renormalized without any
operator mixing between different chiralities.
These facts suggest that the good chiral properties of the physical
Green functions are preserved also at one loop level as in the tree
level.
Furthermore it can be shown that the ${\cal O}(a)$ errors automatically
vanish in the renormalization factors at any loop level in the
perturbation theory \cite{Noaki-Taniguchi}.
These superior features can be understood by the peculiar form of the
external line propagator.
The important point is that the even and odd function $\xi_{e/o}$ is
separated with different damping factor $L_s$ and $R_s$ in the
propagator \eqn{eqn:barq}.

We consider the half-circle diagram of quark self-energy correction
(Fig.~2b of Ref.~\cite{Noaki-Taniguchi}) as an example.
The external line factor $[\xi_e L(p)_s-\xi_o R(p)_s]$,
multiplied by a single gluon interaction vertices $V^{(1)}(p,k)_{st}$,
becomes 
\begin{eqnarray}
[\xi_eL_s-\xi_oR_s]V^{(1)}_{st}= (u_{o}L_t+u_{e}R_t),
\label{dampV1}
\end{eqnarray}
where $u_{\mu e}$ and $u_{\mu o}$ are even and odd function.
In this operation the combination of the even and odd functions with the 
damping factors $L,R$ is flipped,
however the structure that the even and odd function is separated with
$L$ and $R$ is not changed. 
Then this factor is multiplied to the internal fermion propagator
$S_F(p)_{st}$,
\begin{eqnarray}
&&\!\!\!\!\!\!\!\!\!\!\!
(u_{o}L_t+u_{e}R_t)(p)S_F(l)_{st}
\nn\\ &=&\!\!\!\!
 f_o L(p)_t +h_o L(l)_t
+f_e R(p)_t +h_e R(l)_t,
\end{eqnarray}
which does not change the even-odd combination
but only shift the damping ratio.
After being multiplied to another interaction vertex and the
even-oddness being flipped, the factor meets with the damping factor
$[R_s\xi_e-L_s\xi_o]$ from the other external fermion line and
the flavor dependence is summed over,
\begin{eqnarray}
(u_{e}L_s+u_{o}R_s)(R_s\xi_e-L_s\xi_o).
\end{eqnarray}
In this summation only the $L_sL_s$ and $R_sR_s$ combination gives
nonzero even contribution and the result becomes definitely odd function
in terms of the external quark momentum and mass.
The above argument can be applied to other diagrams in any loop level
and we can easily show that the quark self-energy $\Sigma_q$ in
\eqn{eqn:one-loop} is an odd function.
If we expand $\Sigma_q$ in terms of quark momentum and mass keeping the
logarithmic dependence in the coefficients, the leading term and the next 
to next to leading term, which correspond to the additive mass correction 
and the ${\cal O}(a)$ errors, vanish automatically.

Finally we will see how $Z_s =Z_p$ and $Z_V=Z_A$ is obtained
in perturbation theory.
For massless quarks the fermion propagator and the interaction vertex
take the following forms:
\begin{eqnarray}
S_F = \gamma_\mu x_o + x_e
,\quad
V^{(1)} = \gamma_\nu y_e + y_o,
\end{eqnarray}
where $x_e,y_e$ and $x_o,y_o$ are even and odd functions.
Since one loop corrections to bilinear operators
are written as $v_o \Gamma v_o$ with odd functions $v_o$,
$\gamma_5$ in one loop diagrams of the axial vector current and the
pseudo scalar density can be moved outside without changing integrands:
\begin{eqnarray}
\gamma_5 S_F V^{(1)} &\to&
\gamma_5 \left(\gamma_\mu\gamma_\nu x_o y_e+x_e y_o\right)
\nn\\&=& \left(\gamma_\mu\gamma_\nu x_o y_e+x_e y_o\right) \gamma_5 .
\end{eqnarray}
This implies $Z_S=Z_P$ and $Z_V=Z_A$.

%The one loop correction to the conserved axial vector current
%\cite{Shamir95} can be calculated in the same way and gives $Z_A=1$
%\cite{Aoki-Taniguchi}.

%T.~Izubuchi and Y.~Tanigchi are JSPS fellows.

\vspace*{-3mm}

%%%%%%%%%%%%%%%%%%%%%%%%%%%%%%%%%%%%%%%%%%%%%%%%%%%%%%%%%%%%%%%%%%%%%%
\newcommand{\J}[4]{{#1} {\bf #2} (#3) #4}
\newcommand{\AP}{Ann.~Phys.}
\newcommand{\CMP}{Commun.~Math.~Phys.}
\newcommand{\IJMP}{Int.~J.~Mod.~Phys.}
\newcommand{\MPL}{Mod.~Phys.~Lett.}
\newcommand{\NP}{Nucl.~Phys.}
\newcommand{\NPSup}{Nucl.~Phys.~B (Proc.~Suppl.)}
\newcommand{\PL}{Phys.~Lett.}
\newcommand{\PR}{Phys.~Rev.}
\newcommand{\PRL}{Phys.~Rev.~Lett.}
\newcommand{\PTP}{Prog. Theor. Phys.}
\newcommand{\Suppl}{Prog. Theor. Phys. Suppl.}
%%%%%%%%%%%%%%%%%%%%%%%%%%%%%%%%%%%%%%%%%%%%%%%%%%%%%%%%%%%%%%%%%%%%%%

\end{document}